\documentstyle[12pt]{article}

\newcommand{\nl}{\nonumber\\}
\newcommand{\lpar}{\left(}                            
\newcommand{\rpar}{\right)} 
\newcommand{\zb}{Z}
\newcommand{\wb}{W}
\newcommand{\ord}[1]{{\cal O}\lpar#1\rpar}
\newcommand{\mw}{M_{_W}}
\newcommand{\shs}{{\hat s}^2}
\newcommand{\mh}{M_{_H}}
\newcommand{\mt}{m_t}
\newcommand{\mts}{m^2_t}
\newcommand{\mzs}{M^2_{_Z}}
\newcommand{\mz}{M_{_Z}}
\newcommand{\seffsf}[1]{\sin^2\theta^{#1}_{\rm{eff}}}
\newcommand{\gz}{\Gamma_{_Z}}

\newcommand{\als}{\alpha_s}
\newcommand{\gn}{\Gamma_{\nu}}
\newcommand{\gel}{\Gamma_{e}}
\newcommand{\gmu}{\Gamma_{\mu}}

\newcommand{\gt}{\Gamma_{\tau}}

\newcommand{\gu}{\Gamma_{u}}
\newcommand{\gd}{\Gamma_{d}}
\newcommand{\gc}{\Gamma_{c}}
\newcommand{\gs}{\Gamma_{s}}
\newcommand{\gbq}{\Gamma_{b}}
\newcommand{\gh}{\Gamma_{h}}
\newcommand{\gi}{\Gamma_{inv}}
\newcommand{\afba}[1]{A^{#1}_{_{\rm FB}}}
\newcommand{\alra}[1]{A^{#1}_{_{\rm LR}}}

\begin{document}

\bibliographystyle{unsrt}  

\rightline{FNT/T 98/02}
\vspace{0.5cm}

\noindent
{\Large{\bf {\tt TOPAZ0}~4.0 - A new version of a computer program for
evaluation of de-convoluted and realistic observables at LEP~1 and LEP~2}}

\vspace{0.5cm}
\medskip
\noindent
Guido MONTAGNA$^{a,b}$, Oreste NICROSINI$^{b,a}$, 
Fulvio PICCININI$^{b,a}$ \\
\vskip 3pt\noindent
Giampiero PASSARINO$^c$  \\

\noindent
{\it $^a$Dipartimento di Fisica Nucleare e Teorica, Universit\`a di
         Pavia, via A. Bassi n. 6 - 27100 PAVIA - ITALY} \\
\noindent
{\it $^b$ INFN, Sezione di Pavia, via A. Bassi n. 6 -
27100 PAVIA - ITALY} \\
\noindent
{\it $^c$Dipartimento di Fisica Teorica, Universit\`a di Torino, and INFN,
Sezione di Torino, via P. Giuria n. 1 - 10125 TURIN - ITALY} \\

\medskip
\noindent
Program classification: 11.1 \\
\medskip

\hrule \vskip 6pt\noindent

{\small \noindent
{\bf Abstract } \\
The program {\tt TOPAZ0} was developed for computing a variety of
physical observables which are related to the $e^+e^-$ annihilation
into fermion pairs and to the large angle Bhabha scattering around the $\zb$ 
resonance. Among them, the $\zb$ parameters or pseudo-observables, the 
de-convoluted cross sections and those dressed with QED radiation, and
finally the forward-backward asymmetries.
The calculations are performed both for a completely inclusive 
experimental set-up
and for a realistic one, i.e. a set-up with cuts on the acollinearity
angle, on the energy of the outgoing fermions or on their invariant
mass and angular acceptance. The new version, 4.0, includes several
innovative features.
First of all, the most important new capabilities since previous versions are
recently computed electroweak and QCD correction factors that are relevant at 
the $\zb$ resonance in the light of the present experimental accuracy. 
Among them, the effect of the next-to-leading $\ord{\alpha^2\mts}$ 
corrections and those radiative corrections to the 
hadronic decay of the $\zb$ which provide complete corrections of 
$\ord{\alpha\alpha_s}$ to $\Gamma(\zb \to q\bar{q})$ with $q=u,d,s,c$ and $b$.
Secondly, the program has been upgraded to cover two-fermion final states at 
LEP~2 energies, where some of the assumptions made for earlier versions are 
no longer valid. 
In particular, to this aim all the electroweak radiative corrections that are 
negligible at the $\zb$ peak, but relevant far from it, have been added for
$s$-channel processes, 
e.g. purely weak boxes, next-to-leading $\ord{\alpha^2}$ and leading 
$\ord{\alpha^3}$ QED corrections.}

\vskip 6pt  \hrule 

\newpage

\leftline{\Large{\bf NEW VERSION SUMMARY}}
\vskip 15pt

\leftline{{\it Title of program:} {\tt TOPAZ0}~4.0}
\vskip 8pt

\leftline{{\it Catalogue number:} }
\vskip 8pt

\noindent
{\it Program obtainable from:} CPC Program Library, Queen's
University of Belfast, N. Ireland (see application form in this issue); 
also upon request to one of the authors.
\vskip 8pt

\noindent
{\it Reference in CPC for earlier versions of  program:} 
G.~Montagna, O.~Nicrosini,
G.~Passarino, F.~Piccinini and R.~Pittau, Comput. Phys. Commun. 
 76 (1993) 328 (version 1.0); G.~Montagna, O.~Nicrosini,
G.~Passarino and  F.~Piccinini, Comput. Phys. Commun. 
93 (1996) 120 (version 2.0). 
\vskip 8pt

\noindent
{\it Catalogue number of previous version:} ACNT 
\vskip 8pt

\noindent
{\it The new version supersedes the previous ones}
\vskip 8pt

\noindent
\leftline{{\it Licensing provisions:} none}
\vskip 8pt

\noindent
{\it Computers on which this or another recent version has been tested:}
DEC-ALPHA 3000, HP-UX 9000; 
\vskip 8pt

\noindent
\leftline{{\it Operating system under which the new version has been tested:}
VMS, UNIX}
\vskip 8pt

\noindent
{\it Installations:} INFN,
Sezione di Pavia, via A.~Bassi 6, 27100 Pavia, and Sezione di Torino,
via P.~Giuria 1, 10125 Turin, Italy
\vskip 8pt

\noindent
{\it Programming language used:} FORTRAN 77; exception to standard: use 
of {\tt REAL*16} variables
\vskip 8pt

\noindent
{\it Memory required to execute with typical data:} $\approx 300$ kbyte as evaluator 
of observables in seven energy points
\vskip 8pt

\noindent
{\it No. of bits in a word:  32}
\vskip 8pt

\noindent
{\it No. of processors used: } one
\vskip 8pt

\noindent
{\it Subprograms used:} NAGLIB~\cite{nag} 
\vskip 8pt

\noindent
\[ \mbox{\it No. of lines:} \left\{ \begin{array}{rr}
\mbox{PROGRAM} & 15472 \\
\mbox{FORTRAN job-control file} & 180
\end{array}
\right.
\]
 
\vskip 8pt

\noindent
{\it Keywords:} $e^+ e^-$ annihilation, Bhabha scattering, LEP,
$\zb$ resonance, electroweak, extrapolated and realistic experimental set-up,
QCD corrections, QED corrections, pure weak corrections, radiative corrections,
Minimal Standard Model, de-convoluted and realistic observables. 
\vskip 8pt

\noindent
{\it Nature of physical problem} 

\noindent 
An accurate theoretical description of $e^+ e^-$ annihilation
processes and of Bhabha scattering for centre of mass energies 
at the $\zb$ resonance (LEP~1) and above (LEP~2) is necessary in
order to compare theoretical cross sections and asymmetries with the
experimental ones as measured by the LEP collaborations (realistic 
observables). In particular a
{\em realistic } theoretical description, i.e. a description in which the
effects of experimental cuts, such as maximum acollinearity, energy or
invariant mass and angular acceptance of the outgoing fermions, are taken into
account, allows the comparison of the Minimal Standard Model predictions with
experimental {\em raw} data, i.e. data corrected for
detector efficiency but not for acceptance. The program takes
into account all the corrections, pure weak, QED and QCD, which allow for 
such a {\em realistic} theoretical description. 
The program offers also the possibility of computing the $Z$ parameters
(pseudo-observables)
including the state-of-the-art of radiative corrections, which is important for
the indirect determination of the fundamental Standard Model parameters. 
\vskip 8pt

\noindent
{\it Method of solution} 

\noindent
Same as in the original program.
A detailed description of the theoretical formulation and of a sample of
physical results obtained can be found in~\cite{npb93}.
\vskip 8pt

\noindent
{\it Summary of revisions}
\noindent
\begin{itemize}

\item In 1995 the CERN Report on Precision Calculations for the 
$\zb$ resonance~\cite{yr95} provided as basic documentation the 
theoretical basis for 
upgrading existing calculations and FORTRAN programs.

Although the '95 analysis remains quite comprehensive, an update of the 
discussion of radiative corrections has become necessary for one very good
reason: a sizeable amount of theoretical work has appeared following
ref.~\cite{yr95}. In particular, a crucial amount of work has been
performed in providing higher-order corrections.

In ref.~\cite{dfs} the two-loop $\ord{\alpha^2\mts}$ are incorporated in the
theoretical calculation of $\mw$ and $\seffsf{l}$. More recently
the complete calculation of the decay rate of the $\zb$ have been made 
available~\cite{prel}. The only case which is not covered is the one of final 
$b$ quarks, because it involves specific $\ord{\alpha^2\mts}$ vertex corrections.
For simplicity the above correction factors will be referred as 
{\em sub-leadings}.
Another recent development in the computation of radiative corrections to the 
hadronic decay of the $\zb$ is contained in two papers which together provide
complete corrections of $\ord{\alpha\alpha_s}$ to $\Gamma(\zb \to q\bar{q})$
with $q=u,d,s,c$ and $b$.
In the first reference of~\cite{mix} the decay into light quarks is treated. 
In the second one the remaining diagrams contributing to the decay into 
$b$ quarks are considered and thus the mixed two-loops corrections 
are complete.

{\tt TOPAZ0}~\cite{topaz010,topaz020}, 
which was involved in the analysis of refs.~\cite{yr95,yrwg}, 
has been constantly
updated and we focus, in this note, on a short description of the most
important new capabilities since the '95 version.

\begin{itemize}

\item The crucial upgrading refers to the allowed values for {\tt OU0}, 
   a flag contributing to the estimate of the theoretical error. 
   The old option {\tt OU0 = 'Y'/'N'} has now $3$ different entries.
   The effect of the old choice {\tt 'N'} (no re-summation at all of 
   the bosonic corrections, the {\em extreme} option) is disappeared. 
   Instead we have now {\tt OU0 = 'N' or 'L' or 'S'}.
   If one wants to compare with the old default then {\tt OU0 = 'N'}
   (new default) should be selected, {\tt OU0 = 'L'} is a new variant of 
   the re-summation without sub-leadings and could be used to estimate 
   the uncertainty before the introduction of sub-leadings, finally
   {\tt OU0 = 'S'} includes the implementation of the~\cite{dfs}
   correction factors and represents the {\em recommended} choice.
   {\tt TOPAZ0} is based an a series of theoretical options 
   {\tt OUn, n = 0,$\dots$,7}. All these options ($n > 0$) are still active 
   for {\tt OU0 = 'N' or 'L'} but most of them are internally de-activated 
   for {\tt OU0 = 'S'}, thus decreasing the overall uncertainty.
   The default setting remains

\begin{eqnarray*}
{}&{}& {\tt  OU1= 'Y',  OU2= 'N',  OU3= 'Y',  OU4= 'N',}  \nl
{}&{}& {\tt  OU5= 'N',  OU6= 'Y', OU7= 'N'}
\end{eqnarray*}

Changing the parameters is governed by a {\tt FORTRAN} job-control file, 
typical example is: 

\begin{verbatim}
      CHARACTER*1 OVAL
      CHARACTER*3 OFFS
      OFFS= 'OU0'
      OVAL= 'S'
      CALL TCFLAG(OFFS,OVAL)
\end{verbatim}

\item There is a major change with respect to the internal philosophy of 
      {\tt TOPAZ0}. A new subroutine has been introduced, 

\begin{verbatim}
TBASIC(NT,ST,FVECMS,IFLAG) 
\end{verbatim}

      Once {\tt TOPAZ0} is called then the first step is governed by an 
      internal call to subroutine {\tt TWIDHTO}. Here, as a {\em primitive} 
      calculation, both $\mw$ -- the $\wb$ boson mass -- and $\shs$ -- the 
      improved one-loop solution for the sinus 
      of the weak mixing angle -- are obtained as iterative solutions of the 
      two relevant renormalization equations. After that, all $\zb$-parameters 
      are computed. As a matter of fact, most of the users are solely 
      interested in deriving pseudo-observables. For this reason, it is now 
      possible to call {\tt TOPAZ0} with {\tt OEXT = 'P'} where the 
      calculation of the realistic observables is skipped with a considerable 
      gain in CPU time.

\item In computing $\mw, \shs$ and the rest of the pseudo-observables
there are essentially five functions which are touched by the inclusion of
the sub-leadings. Each of these functions depends on $\mh, \mt, \shs$ and
$\mw$. As a function of the ratio $\mh/\mt$ they are given in~\cite{dfs}
in terms of two expansions,$\mh/\mt \gg 1$ and $\mh/\mt \ll 1$. In between it 
is up to the reader to decide what to do. In between, however, is exactly where
the the present experimental data would like to see the Higgs boson. 
In order to avoid any kink in the evaluation of pseudo-observables
one has to interpolate numerically.
The origin of the kink is related to the two-loop correction factor,
$\Delta\rho^{(2)}$, which in the hierarchy of the effects is the dominant one. 
We decided to interpolate with some accuracy for this function. 

\item The most important upgrading in the electroweak/QCD interplay is 
represented by non-factorizable QCD and EW corrections to the hadronic $\zb$ 
boson decay rate.
The Born result receives both QCD and EW corrections and, so far, one used 
factorization
\begin{equation}
\Gamma = \Gamma_{EW}\left( 1+\frac{\alpha_s}{\pi}\right)
\end{equation}
However the correct implementation of the new result~\cite{mix} is giving us
\begin{equation}
\Delta\Gamma\left(\zb \to u,d,s,c\right) \approx -0.59(3)\,{\rm MeV},
\end{equation}
or
\begin{equation}
\Delta\alpha_s(\mz) \approx -\pi\,{{\Delta\Gamma\left(\zb \to 
{\mbox{hadrons}}\right)}\over
{\Gamma\left(\zb \to {\mbox{hadrons}}\right)}} \approx \pi\frac{0.50}{1743} 
\approx 0.001.
\end{equation}

\item There is no space enough here to account for a complete description of 
the QCD corrections implemented in {\tt TOPAZ0}. We simply note that subroutine
{\tt TCORRQCD} has been completely re-designed and function {\tt RRUNM}
allows the running quark masses to be computed at the corresponding pole mass.
For instance subroutine {\tt TCORRQCD} computes

\begin{itemize}
\item[--] the running of the $s$-quark mass up to the $c$- and $b$-quark 
          threshold
\item[--] the $c$-quark mass at the $c$-quark threshold
\item[--] the $c$-quark mass at the $b$-quark threshold
\item[--] and finally the running $c$-quark mass at any scale.
\item[--] The $b$-quark mass at $b$-quark threshold
\item[--] and finally the running $b$-quark mass at any scale.
\end{itemize}

In the following we present a list of the relevant corrections:

\begin{enumerate}

\item The $\ord{\als}$ corrections including non-vanishing quark masses.

\item Flavor non-singlet corrections in the massless limit are the same as 
those for the vector current. Whereas flavor singlet contributions for
vector currents arise only at order $\als^3$ they are present already in
second order for the axial part.

\item Massless non-singlet corrections.

\item Quadratic massive corrections.

\item Quartic massive corrections.

\item Power suppressed $top$-quark mass correction.

\item Singlet axial corrections.

\item Singlet vector correction.

\end{enumerate}

\item It is important to discuss the most relevant variations in the 
prediction for pseudo-observables. We refer to the situation presented
in refs.~\cite{yr95,yrwg}.
\begin{table}[ht]
\begin{center}
\begin{tabular}{|c||c|c|c|c|}
  \hline
  & \multicolumn{4}{c|}{NEW versus OLD} \\
  \cline{2-5}
  & {\tt TOPAZ0 2.0} & {\tt TOPAZ0 4.0} & Rel. shift(per-mil) & Abs. shift \\
  \hline \hline
$\mw\,$[GeV] & 80.310 & 80.308 & -0.03 & -2.1 [MeV]  \\
  \hline \hline
$\seffsf{l}$ & 0.23200 & 0.23209 & 0.39 & 8.9$\times 10^{-5}$  \\
  \hline \hline
$\gz\,$[MeV] & 2497.4 & 2496.1 & -0.052 & -1.29 [MeV]  \\
  \hline \hline
  \hline 
\end{tabular}
\end{center}
\caption[]{Comparison of two TOPAZ0's versions. 
Here $\mz = 91.1888\,$GeV, $\mt = 175\,$GeV, $\alpha_s(\mzs) = 0.125$ and
$\mh = 300\,$GeV.} 
\label{oldnew}
\end{table}
For this reason we have taken again $\mz = 91.1888\,$GeV, $\mt = 175\,$GeV,
$\alpha_s(\mzs) = 0.125$ and $\mh = 300\,$GeV (the '95 input parameter
set) and compared some of the predictions in Table~1. 

\item Purely weak boxes have been introduced for $s$-channel processes. 
Their effect has been already 
discussed at length in~\cite{smp95}.

\end{itemize}

\item QED corrections --- For both the cross sections and forward-backward 
asymmetry, new QED corrections have been added. They include both corrections 
that are relevant at the $\zb$ peak, in view of the present accuracy of 
experimental data, and correction that are negligible at the $\zb$ peak,
but become relevant for centre of mass energies above it.  

\begin{itemize}

\item Cross sections -- New corrections, already known in the literature, 
but not implemented in the previous versions,  and completely new ones have 
been inserted. They are: 

\begin{itemize}

\item initial-state next-to-leading $\ord{\alpha^2}$ corrections to the 
radiator, according to eqs.~(3.19) and (3.20) in~\cite{berends}; this radiator 
is selected by {\tt OHC = Y} and {\tt ORAD = A}; 

\item additional initial-state higher-order corrections to the radiator, 
according to eqs.~(3.29) and (3.30) in~\cite{berends}; this radiator is 
selected by {\tt OHC = Y} and {\tt ORAD = D}; 

\item additional initial-state higher-order corrections to 
the radiator, including part of the leading $\ord{\alpha^3}$ corrections,  
according to eqs.~(3.31) and (3.32) in~\cite{berends}; this radiator is 
selected by {\tt OHC = Y} and {\tt ORAD = E};

\item the full radiator including initial-state next-to-leading 
$\ord{\alpha^2}$ and complete leading $\ord{\alpha^3}$ corrections, according 
to the analytical results obtained in~\cite{a3}; this radiator is selected by 
{\tt OHC = Y} and {\tt ORAD = F} ({\em recommended choice}).  

\end{itemize}

For each of these choices, also the primitives have been computed analytically 
and implemented in {\tt FUNCTION TPRAD}. Moreover, also the 
$\ord{\alpha^3}$ contributions to the electron structure function and its
primitive, according to~\cite{d3}, have been implemented in {\tt SUBROUTINE 
TSTRUCFUN} and {\tt SUBROUTINE TPDFUN}. 

\item Forward-backward asymmetries -- For 
{\tt OHC = Y}, additional contributions to the forward-backward radiator,
that become relevant above the $\zb$ peak, are included, according to 
eqs.~(43), at the $\ord{\alpha}$, and (46), at the $\ord{\alpha^2}$, 
in~\cite{bh}. These contributions are active only for $s$-channel asymmetries. 
Also in this case, the primitive has been computed analytically and 
implemented in {\tt FUNCTION TPRAD}. 

\end{itemize}

{\tt OHC = N} sets automatically the radiator adopted in the previous 
versions of the program, both for cross sections and asymmetries. 

\end{itemize}

The effects of  the above QED corrections have been investigated in~\cite{a3}.
They can be summarized as follows: at LEP~1 the initial-state next-to-leading  
$\ord{\alpha^2}$ corrections are negligible, whereas the leading 
$\ord{\alpha^3}$ ones introduce a negative  shift of the total cross section of
about 0.07\%. At LEP~2, the two corrections tend to compensate one another,
leaving a net effect of around 0.2--0.4\% when the $Z$ radiative return is
included and about 0.1\% otherwise.

It is worth emphasizing, that in order to implement the new radiators and their
primitives, a new library of Nielsen's polylogarithms has been created,

\begin{verbatim}
SUBROUTINE TPOLYL(X,EP,S11,S12,S13,S21,S22)
\end{verbatim}

It returns $S_{1,1}(x), \dots S_{2,2}(x)$.

\vskip 8pt

\noindent
The running is governed by a {\tt FORTRAN} job-control file, 
where the necessary input parameters are fixed. The program calls
{\tt SUBROUTINE TINIT} (see below), where all the flags are initialized to 
their default value. It is allowed to change the default settings by calling 
the auxiliary {\tt SUBROUTINES} {\tt TCUTSET, TCFLAG} and {\tt TCOPT}. 
In particular, the {\em theoretical options} {\tt OUn} are re-set by
subroutine {\tt TCOPT}, kinematical cuts by subroutine {\tt TCUTSET} and 
all the remaining {\tt TOPAZ0}'s flags by subroutine {\tt TCFLAG}.
A change of a non-existing flag is signalled by a warning message:

\begin{verbatim}
TOPFLAG: FLAG NOT RECOGNISED: FLAGNAME
\end{verbatim}

Then {\tt SUBROUTINE TOPAZ0} is called, according to the following calling 
statement:

\begin{verbatim}
SUBROUTINE TOPAZ0(NRTS,RTS,ZMT,TQMT,HMT,ALST,OTPPO,OTPRO). 
\end{verbatim}

The first six entries are input parameters, whose meaning is\\
\noindent
{\tt \underbar{NRTS}}: number of energies;\\
{\tt \underbar{RTS}}: array dimensioned as {\tt RTS(NRTS)}, containing the 
values of the energies;\\
{\tt\underbar{ZMT}}:  the $\zb$-boson mass (GeV);\\
{\tt \underbar{TQMT}}: the $top$-quark mass (GeV);\\
{\tt \underbar{HMT}}: the Higgs-boson mass (GeV);\\
{\tt \underbar{ALST}}:  the value of $\alpha_s(\mzs)$.\\
The last two variables are output quantities:\\
{\tt \underbar{OTPPO}}:  array dimensioned as {\tt OTPPO(24)}, containing the 
values of the pseudo-observables; the internal ordering is:\\
\begin{eqnarray*}
{}\hbox{mass of the W} && \mw(1) \nl
{\hbox{ hadronic peak cross-section}} &&
\sigma_{\rm had}(16) \nl
{\hbox{partial leptonic widths}}&&
 \gn(2), \gel(3), \gmu(4), \gt(5) \nl 
{\hbox{partial hadronic widths}}&&
\gu(6), \gd(7), \gc(8), \gs(7), \gbq(9)  \nl
{}\hbox{the total width} &&     \gz(14)  \nl
{\hbox{the total hadronic width}} && \gh(19)  \nl
\hbox{the total invisible width} && \gi(20) \nl
{\hbox{ratios}}&& R_l(15), R_b(17), R_c(22) \\
{\hbox{asymmetries and polarization}}&&
 \afba{l}(12), \alra{l}(13), \afba{b}(18), \nl
{}&{}& \afba{c}(21), \alra{b}(23), \alra{c}(24) \\
{\hbox{{effective sinuses}}}&& \seffsf{l}(10), \seffsf{b}(11) \\
\end{eqnarray*}
\noindent
{\tt \underbar{OTPRO}:}  array dimensioned as {\tt OTPRO(26*NRTS)}, containing 
the values of the realistic observables and relative numerical errors:\\
\[
{\tt RO} = \left\{ \begin{array}{llll}
\sigma_e \pm \Delta \sigma_e &  
\sigma_{\mu} \pm \Delta \sigma_{\mu} &
\sigma_{\tau} \pm \Delta \sigma_{\tau} & \\
\sigma_{\rm had} \pm \Delta \sigma_{\rm had} & & &  \\
\sigma_b \pm \Delta \sigma_b &        
R_b \pm \Delta R_b & 
\sigma_c \pm \Delta \sigma_c &        
R_c \pm \Delta R_c  \\
\afba{e} \pm \Delta \afba{e} &        
\afba{\mu} \pm \Delta \afba{\mu} &
\afba{\tau} \pm \Delta \afba{\tau} & \\
\afba{c} \pm \Delta \afba{c} &
\afba{b} \pm \Delta \afba{b} & &
\end{array}
\right.
\]
Results are stored in array {\tt OTPRO(K)} according to {\tt 
K = 13\,(2(I-1)+J-1)+L} with {\tt I = 1,NRTS} running over energy points, 
{\tt L = 1,13} running over the type of realistic observable ($\sigma_e, 
\sigma_{\mu}\,$ etc.) and {\tt J = 1,2} for the central value and the 
numerical error. 

The output is still governed by the job-control file, 
by means of the printing of the arrays {\tt OTPPO(24)} and 
{\tt OTPRO(26*NRTS)}, to be declared in {\tt COMMON}. The output from the 
program is self-explanatory. 

All the internal flags of the program are initialized by means of 
{\tt SUBROUTINE TINIT}, which is called by the job-control file.  
Here the complete list of inputs is given, devoting particular care in 
commenting the meaning of the new flags. 
For more details concerning flags already present in earlier versions, the 
reader is referred to the documentation of the previous releases. \\
{\tt \underbar{SE}}: scaling factor for numerical integration error;\\
{\tt \underbar{OWEAK}}: residual weak corrections: running {\tt (R)} or fixed 
{\tt (F)}; at LEP~2 the correct  choice is {\tt R}; \\
{\tt \underbar{OWBOX}}: weak boxes included {\tt (Y)} or not {\tt (N)}.
Away from the $\zb$-resonance the {\em correct} choice is {\tt Y};\\
{\tt \underbar{OU0}}: includes {\tt (S)} the sub-leading two-loop 
$\ord{\alpha^2\mts}$ corrections according to~\cite{dfs}. It is the
{\em recommended} choice; \\
{\tt \underbar{OHC}}: next-to-leading and higher order hard photon 
contribution included {\tt (Y)} or not {\tt (N)};\\
{\tt \underbar{ORAD}}: selects the type of next-to-leading and higher order 
hard photon contributions to be included {\tt (A,D,E,F)}. {\tt F} is the
{\em recommended} choice;\\
{\tt \underbar{OAL}}: selects the value of $\alpha_{em}^{-1}(\mzs)$: 
128.896 {\tt (Y)}, 128.87 {\tt (N)}, user-defined {\tt (V)}; if
{\tt OAL = 'V'} then a new value for $\alpha_{em}^{-1}(\mzs)$ is selected
through the calling procedure {\tt CALL TCAQED(ALPHANEW)}.\\
{\tt \underbar{OAAS}}: the scale in $\alpha_s$ in mixed $\alpha \alpha_s$ 
corrections;\\
{\tt \underbar{ONP}}: array dimensioned as {\tt ONPT(NRTS)} 
selecting {\tt (Y)} or not {\tt (N)} the pair production correction for each 
energy;\\
{\tt \underbar{ONIF}}: array dimensioned as {\tt ONIF(NRTS)} 
selecting {\tt (Y)} or not {\tt (N)} the effect of initial-finale state 
QED interference for each energy;\\
{\tt \underbar{OEXT}}: selects the extrapolated {\tt (E)} or cut {\tt (C)} 
branch for $\mu$ and $\tau$. With {\tt OEXT = 'P'} pseudo-observables are 
computed and control is returned;\\
{\tt \underbar{OFB}}: selects {\tt (Y)} or not {\tt (N)} different treatment 
of cuts for cross sections and forward-backward asymmetries for $\mu$ and 
$\tau$;\\
{\tt \underbar{OCUTS}}: array dimensioned as {\tt OCUTST(NRTS)}; if 
{\tt OFB = Y} selects {\tt (Y)} or not {\tt (N)} a cut on the 
invariant mass after initial-state radiation for $\mu$ and $\tau$ cross 
sections; \\
{\tt \underbar{OCHAN}}: selects the full Bhabha scattering matrix element 
{\tt (F)} ({\em default}) or only the $s$-channel part {\tt (S)};\\
{\tt \underbar{OCUT}}: array dimensioned as {\tt OCUT(NRTS)}; selects a cut 
on $s'$: no cut {\tt (NC)}, cut on hadronic channels only {\tt (HC)}, cut on 
all channels {\tt (FC)}. Here $s'$ is the invariant mass of the $e^+e^-$ 
system after initial state radiation;\\
{\tt \underbar{OCUTF}}: as {\tt OCUT}, but for final-state fermions invariant 
mass;\\
{\tt \underbar{OTHRMT}}: if {\tt OEXT = C}, selects invariant mass {\tt (M)} 
or energy {\tt (E)} threshold for $\mu$ and $\tau$; \\
{\tt \underbar{OCREE}}: if {\tt OEXT = E} and {\tt OCHAN = S}, i.e. for 
$s$-channel electrons, it allows for an $s'$ cut; \\
{\tt \underbar{OCUTES}}: array dimensioned as {\tt OCUTES(NRTS)}; if 
{\tt OCREE = Y}, it allows {\tt (C)} or not {\tt (N)} for an $s'$ 
cut for each energy point;\\
{\tt \underbar{OTHRE}}: selects invariant mass {\tt (M)} or energy {\tt (E)} 
threshold for electrons;\\
{\tt \underbar{OFS}}: selects the treatment of higher order final-state QED 
corrections; {\em default} {\tt (D)} or perturbative {\tt (Z)};\\
{\tt \underbar{OCN}}: includes {\tt (Y)} or not {\tt (N)} the contribution 
of electromagnetic jets for final-state calorimetric electrons;\\
{\tt \underbar{XMED}}: fictitious separator for 1-dimensional integrations 
to improve numerical convergence;\\
{\tt \underbar{OVNAL}}: if {\tt OAL = V}, enters the user-defined value 
$\alpha_{em}^{-1} (\mzs)$ = {\tt OVNAL};\\
{\tt \underbar{SC}}: if {\tt OAAS = Y}, defines the scale in $\alpha_s$ in 
mixed $\alpha \alpha_s$ corrections;\\
{\tt \underbar{ZPCUT}}: the minimum fraction of squared invariant mass 
of the final state after radiation of the additional initial-state pair; \\
{\tt \underbar{OXCUTS}}: if {\tt OCUTS = Y}, it enters the value of the $s'$ 
cut, $s_0 / s$;\\
{\tt \underbar{OXCUT}}: if {\tt OCUT = Y}, it enters the value of the $s'$ 
cut, $s_0 / s$;\\
{\tt \underbar{OXCUTF}}: if {\tt OCUTF = Y}, it enters the cut on the value 
of invariant mass squared of the final-state fermions;\\
{\tt \underbar{S0CUT}}: array dimensioned as {\tt S0CUT(3)}; {\tt S0CUT(1)} 
is the minimum invariant mass (GeV) of the final-state electrons, used if 
{\tt OTHRE = M};  {\tt S0CUT(2)} and {\tt S0CUT(3)} are the same as 
{\tt S0CUT(1)} for $\mu$ and $\tau$ respectively, used if {\tt OTHRMT = M}; \\
{\tt \underbar{E0}}: array dimensioned as {\tt E0(3)}; {\tt E0(1)} is the 
minimum energy (GeV) of the final-state electrons, used if {\tt OTHRE = E};  
{\tt E0(2)} and {\tt E0(3)} are the same as {\tt E0(1)} for $\mu$ and $\tau$ 
respectively, used if {\tt OTHRMT = E}; \\
{\tt \underbar{THMIN}}: array dimensioned as {\tt THMIN(3)}; it enters the 
minimum scattering angle (deg) for electrons, $\mu$ and $\tau$, respectively 
({\em symmetrical angular acceptance} is understood); {\tt THMIN(2)} and 
{\tt THMIN(3)} used only if  {\tt OEXT = C}; \\
{\tt \underbar{THMINP}}: the same as {\tt THMIN}, for the antiparticles;\\
{\tt \underbar{ACOLL}}: array dimensioned as {\tt ACOLL(3)}; it enters the 
maximum acollinearity angle (deg) for electrons, $\mu$ and $\tau$, 
respectively; {\tt ACOLL(2)} and {\tt ACOLL(3)} used only if  {\tt OEXT = C}; \\
{\tt \underbar{OXCUTES}}: if {\tt OCUTES(I) = C}, it enters the $s_0 / s$ 
threshold for the I-th energy point;\\
{\tt \underbar{DEL}}: the semi-aperture of the electromagnetic jet (deg); 
used only if {\tt OCN = 'Y'}.\\

The internal default has been set to the following choices:

\begin{eqnarray*}
&& {\tt OWEAK= 'R',   OWBOX= 'Y', OHC= 'Y',    ORAD= 'F'}  \nl
&& {\tt XMED= 0.98D0, OAL= 'Y',   OCN= 'Y',    DEL= 0.5D0}  \nl
&& {\tt ZPCUT= 0.8D0, OFS= 'D',   ONP(I)= 'Y', ONIF(I)= 'Y'}
\end{eqnarray*}

A change for some of the kinematical cuts is illustrated by the following
calling sequence:

\begin{verbatim}
      IND= 1
      DMY= -1.D0
      ANG= 40.D0
      ACL= 10.D0
      ETH= 5.D0
      CALL TCUTSET(IND,DMY,ANG,DMY,ACL,ETH,DMY,DMY,DMY)
\end{verbatim}

With {\tt IND = 1} only cuts for the electron will be touched and moreover
{\tt DMY = -1}, i.e. DMY$\,<\, 0$ implies that variables {\tt S0CUT(1), 
THMINP(1), OXCUT, OXCUTF} and {\tt ZPCUT} keep the default value, while 
variables {\tt THMIN(1), ACOLL(1)} and {\tt E0(1)} are reset to new
values $40^\circ, 10^\circ$ and $5\,$GeV.

After initialization a call is performed to subroutine {\tt TWIDTHO} after
which the control is returned if {\tt OEXT = 'P'}, otherwise 
subroutine {\tt TEWEXT} and {\tt TEWCUT} are called and the evaluation
of realistic observables starts. In {\tt TEWEXT} weak corrections are computed 
and the physical quantities are convoluted when no cuts (but on the invariant 
mass) are applied. 
In {\tt TWCUT} weak corrections are computed and the physical quantities are
convoluted for leptons, when cuts are applied. 

The previous versions of {\tt TOPAZ0} can be used to compute $\zb$-parameters 
and de-convoluted observables but also to obtain predictions for QED-dressed 
distributions, over some realistic set-up, resembling the experimental 
{\it raw} data. 
The most severe limitation of those versions is that they have been developed 
by having in mind the accurate evaluation of theoretical observables at the 
$\zb$ peak. 
The new version of the program, {\tt TOPAZ0}~4.0, offers remarkable 
improvements in two directions. First, towards
the goal of taking into account all those radiative corrections that can 
become relevant, when we consider the experimental accuracy reached at 
present by the LEP~1 experiments. Of the same relevance is the effort 
of adding those radiative corrections that are negligible for 
a centre-of-mass energy around the $\zb$ peak, but that become relevant for 
energies above it.   

\vskip 15pt
\noindent
{\it Restrictions on the complexity of the problem} 

\noindent
Analytic formulas have been developed for an experimental set-up with
symmetrical angular acceptance. Moreover the angular acceptance of the
scattered antifermion has been assumed to be larger than the one of the 
scattered fermion. The prediction for Bhabha scattering is understood to 
be for the large-angle regime.
Initial-state next-to-leading $\ord\alpha$ QED corrections are treated
exactly for an $s'$ cut, 
in the soft photon
approximation otherwise. 
This means that for center of mass energies sensibly above 
the $\zb^0$ peak (typically in the LEP~1.5 -- LEP~2 regime), the theoretical 
accuracy of the {\tt C} branch is under control  (theoretical error $\le$ 
0.3\%) when excluding the $\zb$ radiative return, whereas including it the 
theoretical error can grow up to some \% depending on the final state 
selected~\cite{ishp}. In the same energy range, large angle Bhabha scattering 
becomes a $t$-channel dominated process: since all the QED corrections 
implemented are strictly valid for $s$-channel processes, this means that 
large angle Bhabha scattering off the $\zb$ resonance is treated at the 
leading logarithmic level.  
\vskip 8pt

\noindent
{\it Typical running time} 

\noindent
This depends strongly on the particular experimental set-up studied and on the
energy range. As evaluator of realistic observables in seven energy 
points around the $Z^0$ peak, between 10 
(extrapolated set-up) and 270 (realistic set-up) CPU seconds for HP-UX 9000. 
Anyway, for the realistic observables the CPU time depends strongly on being at
LEP~1 or LEP~2, and on the scaling factor {\tt SE} controlling the 
accuracy of numerical integrations. For the evaluation of pseudo-observables
the program runs much faster. 

\vskip 8pt

\noindent
{\it Unusual features of the program} 

\noindent
Subroutines from the library of mathematical subprograms NAGLIB~\cite{nag}
for the numerical integrations are used in the program.
\vskip 8pt

\noindent
{\it Acknowledgements}

\noindent
We would like to express special thanks to many colleagues.
Without their contributions the program would not be what it is now.
We have received a great support from the experimental community, but
among them we take the pleasure to acknowledge the active role of
Robert Clare, Peter Clarke, Manel Martinez, Alexander Olshevsky and 
Frederic Teubert. 
Essential, in the development of {\tt TOPAZ0}, 
has been the constant support of 
Martin Gruenewald, Christoph Pauss and Gunter Quast. 
Among our theorist friends we would like to thank Wim Beenakker, Wolfgang 
Hollik, Hans K\"uhn and Roberto Pittau.
We acknowledge the important role played by Giuseppe Degrassi and by
Paolo Gambino in helping us with the implementation of the two-loop
sub-leadings corrections and for sharing with us the result of their work
prior to publication.
Finally we express our gratitude to Dmitri Bardin. It is only due to the 
constant exchange of information and to the continuous cross-checks between 
{\tt TOPAZ0} and {\tt ZFITTER} 
that we have reached the present level of confidence in 
our results.
Perhaps it is not that frequent that competitors forget about competition 
and sit down until they understand everything about their respective work
before publication of the work itself.

\bibliography{topaz0_40}

\begin{thebibliography}{10}

\bibitem{nag}
NAG Fortran Library Manual Mark 17 (Numerical Algorithms Group, Oxford, 1991).

\bibitem{npb93}
G.~Montagna, O.~Nicrosini, G.~Passarino, F.~Piccinini and R.~Pittau,
  Nucl.~Phys. B401 (1993) 3.

\bibitem{yr95}
{\it ``Reports of the Working Group on Precision Calculations for the Z
  Resonance''}, CERN Report 95-03 (Geneva, 1995), edited by D.~Bardin,
  W.~Hollik and G.~Passarino.

\bibitem{dfs}
G.\,Degrassi, S. Fanchiotti, A. Sirlin, Nucl.Phys. B351 (1991) 49; \\ G.
  Degrassi, A. Sirlin, Nucl. Phys. B352 (1991) 342; \\ G. Degrassi, P. Gambino,
  and A. Sirlin, Phys. Lett. B394 (1997) 188; \\ G. Degrassi, P. Gambino, and
  A. Vicini, Phys. Lett. B383 (1996) 219.

\bibitem{prel}
G. Degrassi and P. Gambino, in preparation.

\bibitem{mix}
A. Czarnecki and J.H. K\"uhn, Phys. Rev. Lett. 77 (1996) 3955 and
  hep-ph/9712228; \\ R. Harlander, T. Seidensticker and M. Steinhauser,
  hep-ph/9712228.

\bibitem{topaz010}
G.~Montagna, O.~Nicrosini, G.~Passarino, F.~Piccinini and R.~Pittau, Comput.
  Phys. Commun. 76 (1993) 328.

\bibitem{topaz020}
G.~Montagna, O.~Nicrosini, G.~Passarino and F.~Piccinini, Comput. Phys. Commun.
  93 (1996) 120.

\bibitem{yrwg}
D.~Bardin et al., {\it ``Electroweak Working Group Report''}, in~\cite{yr95},
  p.~7, {\tt hep-ph/9709229}.

\bibitem{smp95}
F.~Boudjema, B.~Mele et al., ``Standard Model Processes'', in {\it Physics at
  LEP2}, G.~Altarelli, T.~Sj\"ostrand and F.~Zwirner, eds., CERN Report {
  96-01} (Geneva, 1996), vol.~2, p.~229.

\bibitem{berends}
F.~Berends et al., ``$Z$ line shape'', in {\it $Z$ Physics at LEP 1},
  G.~Altarelli, R.~Kleiss and C.~Verzegnassi, eds., CERN Report { 89-08}
  (Geneva, 1989), vol.~1, p.~89.

\bibitem{a3}
G. Montagna, O. Nicrosini and F. Piccinini, Phys.~Lett. { B406} (1997) 243.

\bibitem{d3}
M.~Cacciari, A.~Deandrea, G.~Montagna and O.~Nicrosini, Europhys. Lett. { 17}
  (1992) 123.

\bibitem{bh}
M.~B\"ohm, W.~Hollik et al., ``Forward-Backward Asymmetries'', in {\it $Z$
  Physics at LEP 1}, G.~Altarelli, R.~Kleiss and C.~Verzegnassi, eds., CERN
  Report { 89-08} (Geneva, 1989), vol.~1, p.~203.

\bibitem{ishp}
G. Montagna, O. Nicrosini and F. Piccinini, Z.~Phys.~C 76 (1997) 45.

\end{thebibliography}

\newpage

\leftline{\Large\bf  Test Run Output}
\vskip 10pt

The typical calculations that can be performed with the new version of the
program are illustrated in the following example. The FORTRAN job-control file
computes both pseudo-observables and realistic observables with 
{\tt OEXT = 'E'}. For the large angle Bhabha observables we use the following
selection criterion:

\begin{itemize}

\item[--] minimum scattering angle for the $e^-$, $|\cos\theta_e^-| < 0.7$;
\item[--] maximum acollinearity, $\theta_{\rm acol} = 10^\circ$;
\item[--] energy thresholds, $E(e^{\pm}) = 1\,$GeV.

\end{itemize}  

The realistic observables are computed for three typical energies, $\sqrt{s} =
\mz, 136.22\,$GeV and $172.12\,$GeV. The input parameter set is
$\mz = 91.1867\,$GeV, $\mt = 175.6\,$GeV, $\mh = 300\,$GeV and
$\alpha_s(\mzs) = 0.120$.

\begin{verbatim}

 W MASS   (GEV)        =    0.803121E+02
 NU                    =    0.167123E+00
 ELECTRON              =    0.839180E-01
 MUON                  =    0.839173E-01
 TAU                   =    0.837263E-01
 UP                    =    0.299863E+00
 DOWN(STRANGE)         =    0.382626E+00
 CHARM                 =    0.299806E+00
 BOTTOM                =    0.375542E+00
 SIN^2(E)              =    0.232068E+00
 SIN^2(B)              =    0.233372E+00
 A_FB(L) EFF.          =    0.152848E-01
 A_LR EFF.             =    0.142757E+00
 TOTAL WIDTH (GEV)     =    0.249338E+01
 G_H/G_E               =    0.207399E+02
 SIGMA0_H  (NB)        =    0.414744E+02
 G(B)/G(HAD)           =    0.215772E+00
 A_FB(B)               =    0.999354E-01
 HADRONIC WIDTH (GEV)  =    0.174045E+01
 INVISIBLE             =    0.501370E+00
 A_FB(C)               =    0.713261E-01
 G(C)/G(HAD)           =    0.172258E+00
 A_LR(B) EFF.          =    0.934328E+00
 A_LR(C) EFF.          =    0.665952E+00

 E_CM (GEV)            =     0.91187E+02

 SIGMA(E)   (NB)       =   0.1016832E+01 +/-   0.6896812E-05
 SIGMA(MU)  (NB)       =   0.1471494E+01 +/-   0.4361508E-07
 SIGMA(TAU) (NB)       =   0.1468142E+01 +/-   0.4351378E-07
 SIGMA(HAD) (NB)       =   0.3034906E+02 +/-   0.3921926E-06
 SIGMA(B)   (NB)       =   0.6543238E+01 +/-   0.1888141E-06
 R_B                   =   0.2155994E+00 +/-   0.6816784E-08
 SIGMA(C)   (NB)       =   0.5235138E+01 +/-   0.1483041E-06
 R_C                   =   0.1724975E+00 +/-   0.5371037E-08
 A_FB(E)               =   0.1074445E+00 +/-   0.7509969E-05
 A_FB(MU)              =   0.1597633E-03 +/-   0.9067817E-07
 A_FB(TAU)             =   0.1612020E-03 +/-   0.9074563E-07
 A_FB(C)               =   0.5966220E-01 +/-   0.5177077E-07
 A_FB(B)               =   0.9418619E-01 +/-   0.2164750E-07

 E_CM (GEV)            =     0.13622E+03

 SIGMA(E)   (NB)       =   0.3910009E-01 +/-   0.3801224E-05
 SIGMA(MU)  (NB)       =   0.7110528E-02 +/-   0.9921073E-07
 SIGMA(TAU) (NB)       =   0.7108396E-02 +/-   0.9910925E-07
 SIGMA(HAD) (NB)       =   0.6284555E-01 +/-   0.9266594E-06
 SIGMA(B)   (NB)       =   0.1137215E-01 +/-   0.4471414E-06
 R_B                   =   0.1809539E+00 +/-   0.7598769E-05
 SIGMA(C)   (NB)       =   0.1427977E-01 +/-   0.3536671E-06
 R_C                   =   0.2272201E+00 +/-   0.6549381E-05
 A_FB(E)               =   0.8019347E+00 +/-   0.1750868E-03
 A_FB(MU)              =   0.6970068E+00 +/-   0.1225792E-04
 A_FB(TAU)             =   0.6967559E+00 +/-   0.1224640E-04
 A_FB(C)               =   0.6910530E+00 +/-   0.2064462E-04
 A_FB(B)               =   0.4956733E+00 +/-   0.2387142E-04

 E_CM (GEV)            =     0.17212E+03

 SIGMA(E)   (NB)       =   0.2439700E-01 +/-   0.3792062E-05
 SIGMA(MU)  (NB)       =   0.3878666E-02 +/-   0.2972920E-07
 SIGMA(TAU) (NB)       =   0.3878186E-02 +/-   0.2971010E-07
 SIGMA(HAD) (NB)       =   0.2754651E-01 +/-   0.2768351E-06
 SIGMA(B)   (NB)       =   0.4584642E-02 +/-   0.1347729E-06
 R_B                   =   0.1664328E+00 +/-   0.5170563E-05
 SIGMA(C)   (NB)       =   0.6889562E-02 +/-   0.1053195E-06
 R_C                   =   0.2501065E+00 +/-   0.4575542E-05
 A_FB(E)               =   0.8091031E+00 +/-   0.2815138E-03
 A_FB(MU)              =   0.6097632E+00 +/-   0.5588466E-05
 A_FB(TAU)             =   0.6095878E+00 +/-   0.5584414E-05
 A_FB(C)               =   0.6717413E+00 +/-   0.1127397E-04
 A_FB(B)               =   0.5592299E+00 +/-   0.1854536E-04


\end{verbatim}

\end{document}